 \documentclass[final,5p,times,twocolumn,authoryear]{elsarticle}
 
\usepackage{tikz-feynman}
\usepackage{amssymb}
\usepackage{lipsum}
\usepackage{cuted}
\usepackage{natbib}
\setcitestyle{numbers,square}

\usepackage{hyperref} 
\usepackage{nicefrac}
\usepackage{amsmath,amsfonts,amsthm,amssymb}
\usepackage{multirow}
\usepackage{multicol}
\usepackage{nicefrac}
\usepackage{cleveref}
\usepackage{yfonts}
\usepackage{slashed}
\hypersetup{
colorlinks = true,
linkcolor = blue,
anchorcolor = blue,
citecolor = blue,
filecolor = blue,
urlcolor = blue
}
\usepackage{ulem}

\journal{Physics Letters B}

\begin{document}

\begin{frontmatter}
\title{Mass spectrum of $S$-wave mesons in the relativistic independent quark model}

\author[first]{Lopamudra Nayak\corref{cor1}}
\ead{lopalmnayak@niser.ac.in}
\affiliation[first]{organization={National Institute of Science Education and Research, An OCC of Homi Bhabha National Institute},
            addressline={Jatni}, 
            city={Bhubaneswar},
            postcode={752050}, 
            state={Odisha},
            country={India}}
\author[first]{Sonali Patnaik} 
\author[first]{Divyajyoti Pandey}
\author[first]{Sanjay Kumar Swain} 

\cortext[cor1]{Corresponding author}

\begin{abstract}
The confining strength or model parameters and constituent quark masses are reparametrized for predicting the ground state meson masses. We analyzed these from the hyperfine splitting of $S$-wave heavy-flavored, heavy-light and light (except non-strange) mesons in the framework of a relativistic independent quark (RIQ) model with one gluon exchange and centre-of-mass correction. These mesons with spin parity $J^P=0^-$ and $1^-$, the masses obtained are in accordance with the experimental physical masses. The results will serve as good complementary tools in further study of hadron dynamics and will behave as a foundation for the higher excited and exotic states of hadrons.  
\end{abstract}




\end{frontmatter}




\section{Introduction}
\label{introduction}
In view of the wealth of experimental data \cite{ParticleDataGroup:2024cfk}, hadron spectroscopy could constitute a good testing ground for non-perturbative Quantum Chromodynamics (QCD), as it best for the full meson spectrum, including heavy-light mesons and also exotic states. Since the exact form of confinement from QCD is not known, one has to go for phenomenological models. The phenomenological models are either non-relativistic quark models or relativistic models where the interaction is treated perturbatively. Non-relativistic potential models~\cite{Shlyapnikov:2000iw,Bhavyashri:2005rb,Li:2019tbn} inspired by the basic QCD characteristics have been quite successful for the prediction of various meson properties with the help of the Schr$\ddot{o}$dinger equation.  Masses of the ground as well as excited states can be evaluated by making use of the one gluon exchange (OGE) potential, while the decay properties can be evaluated by using the wavefunctions and the overlap integral. But it ignores the relativistic correction as a result it fails to explain in the light-sector. In contrast, relativistic quark models involving Bathe-salpeter equation and Dyson-Schwinger equation \cite{Chen:2020ecu,Fischer:2014cfa,Ebert:2009ua,Godfrey:2004ya,Solovev:1997htw} have been widely used to study both heavy and light mesons, incorporating relativistic effects like spin-dependent interactions and dynamical chiral symmetry breaking. Furthermore, lattice QCD \cite{Davies:1994mp,Davies:1998im,Thomas:2017rkj} provides a non-perturbative, first-principles approach to studying meson properties, effectively capturing these relativistic effects, providing highly accurate numerical predictions, but requires extensive computational resources. Hadronic mass spectra play a vital role in uncovering the forces that ensure the permanent confinement of quarks and antiquarks, preventing their direct observation as free particles. Empirical evidence from meson hyperfine splitting confirms that the short-range component of the binding potential behaves as expected from single-color gluon exchange in an asymptotically free theory. Since the interaction of light confined quarks is governed by a single length scale, whereas heavy confined quarks interact over two distinct length scales, spin-dependent interactions are expected to differ qualitatively between these cases. Phenomenological studies indicate that spin-dependent interactions among light quarks are predominantly short-ranged, while those among heavy quarks exhibit a primarily long-range character. To further investigate the possibility of a long-range spin-spin interaction in heavy quark systems, it is suggested that measuring the $M^*-M$ mass difference could provide valuable insights.\\
  
Without assuming any specific theoretical mechanism for quark confinement, we adopt an alternative yet comparable approach based on a purely phenomenological individual quark potential in an equally mixed scalar-vector harmonic form. This potential model has been successfully applied in our previous studies to reasonably predict the core contributions to the magnetic moments of octet baryons, the proton charge radius, and the weak electric and magnetic form factors for semileptonic baryon and meson decays~\cite{Barik:1985in,Barik:1985rm}. The model has also been recently employed to study various decays~\cite{Nayak:2021djn,Nayak:2022qaq,Patnaik:2023ins,Nayak:2024esq}. In the present work, we employ a chiral potential model to study the mass spectrum of  heavy-flavored, heavy-light and light (except non-strange) mesons by taking into account the corrections due to the energy associated with the centre-of-mass motion, and the residual-one gluon exchange interaction which includes color-electric and color-magnetic energy. We treat all these corrections ultimately leading to the masses of mesons. This model, specifically employing a harmonic form for the scalar-vector mixed potential, proves to be both simple and tractable in these aspects. It provides highly satisfactory results, not only for the physical masses of the heavy~\cite{Barik:1993aw} and light~\cite{Barik:1987zb} mesons but also for low-lying baryons~\cite{Barik:1986mq}. In this work, we improved the previously utilized potential parameters by fixing the quark masses and adjusting the strong coupling constant according to the energy scale. This refinement leads to an improved agreement with the experimentally observed meson masses. The proposed parameterization may also provide stronger evidence for studies of decay dynamics.

 The manuscript is organized as follows. Following this introduction, a brief overview of the theoretical framework of the potential model is provided in Sec.~\ref{sec:TF} with solutions for the relativistic bound states of the individually confined quarks in the ground state of meson. Sec.~\ref{sec:Res} is mainly dedicated to the analysis and discussion of our theoretical results. Finally in Sec.~\ref{sec:Sum&con} we summarize our work with a concluding remark.
 
\section{Theoretical Formalism}
\label{sec:TF}
In the present model, meson is an assembly of a quark and an antiquark with appropriate interactions according to QCD. The quark-gluon interaction originating from one-gluon exchange at short distances and the quark-pion interaction in the non-strange flavor sector~\cite{Barik:1987zb} required to preserve chiral symmetry are presumed to be residual interactions compared to the dominant confining interaction. To a first approximation, the confining part of the interaction is believed to provide the fundamental quark dynamics within the mesonic core, determining the zeroth-order energy of the  $(q_i\bar{q}_j)$-system $(i,j=u,d,s,c,b)$ (except non-strange light mesons). Since the confining interaction is believed to arise from a non-perturbative multi-gluon mechanism, it remains theoretically intractable to compute directly from first principles. Consequently, from a phenomenological perspective, the current model assumes that quarks within a mesonic core experience independent confinement through an average, flavor-independent potential of the form
\begin{equation}
    U(r)=\frac{1}{2}(1+\gamma^0)(ar^2+V_0)
    \label{eq:pot}
\end{equation}
with $a>0$. Here, $(a, V_0)$ are the
potential parameters. This confining interaction is believed to provide phenomenologically the zeroth order quark dynamics inside the hadron-core through the quark Lagrangian density 
\begin{equation}
{\cal L}^{0}_{q}(x)={\bar \psi}_{q}(x)\;[\;{i\over 2}\gamma ^{\mu}
\partial _{\mu}-m_{q}-U(r)\;]\;\psi _{q}(x)
\label{ld}
\end{equation}
Considering all the quarks in a meson-core in their ground state, the normalized quark wave function $\Psi_{q^+}(r)$ satisfying the Dirac equation, obtainable from ${\cal L}^{0}_{q}(x)$ can be written in the two component form as, 
\begin{eqnarray}
\Psi_{q^+}(\vec r)\;&=&\;{1\over \sqrt{4\pi}}\left(
\begin{array}{c}
ig_{q}(r)/r\\
({\vec \sigma}.{\hat r})f_{q}(r)/r
\end{array}\;\right){\chi _{\uparrow}}
\label{eq:psiq}
\end{eqnarray}
and
\begin{eqnarray}
\Psi_{q^-}(\vec r)\;&=&\;{1\over \sqrt{4\pi}}\left(
\begin{array}{c}
i({\vec \sigma}.{\hat r})f_{q}(r)/r\\
g_{q}(r)/r
\end{array}\;\right)\tilde{\chi _{\downarrow}}
\label{eq:psibarq}
\end{eqnarray}
for positive and negative energy solutions respectively. \\
Taking, 
\begin{eqnarray}
    E'_q=(E_q-V_0/2),\ m'_q=(m_q+V_0/2)\\
    \lambda_q=(E'_q+m'_q) \ \ \text{ and } r_{0q}=(a\lambda_q)^{1/4}
    \label{eq:eqlq}
\end{eqnarray}
it can be shown that the reduced radial parts of the upper and lower component of
$\Psi_{q^+}(r)$ come out as,
\begin{eqnarray}
    g_q(r) =&& N_q(r/r_{0q})exp(-r^2/2r_{0q}^2),\\
f_q(r)=&& -\frac{N_q}{\lambda_q r_{0q}}(r/r_{0q})^2 exp(-r^2/2r^2_{0q})
\label{eq:gqfq}
\end{eqnarray}
The overall normalization factor $N_q$ is given by the relation,
\begin{equation}
    N_q^2=\frac{8\lambda_q}{\sqrt{\pi}r_{0q}}\frac{1}{(3E'_q+m'_q)}
\end{equation}
The energy eigen value condition required to obtain the value of binding energy $E_q=(E'_q + V_0/2)$ of individual quark in ground-state is
\begin{equation}
    ({E'_q}^2 -{m'}^2_q)r^2_{0q}=3
\end{equation}
The solutions through equations \eqref{eq:psiq}--\eqref{eq:gqfq} provide the quark binding energy $E_q$ which immediately leads to the energy of the meson-core in zeroth order as,
\begin{equation}
    E_M^0=\sum_q E_q
    \label{eq:ezerom}
\end{equation} 
\subsection{One gluon-exchange correction}
\label{subsec:oge}
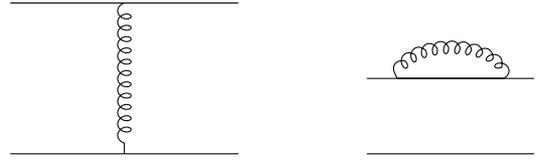
\begin{figure}[hbt]
    \centering
    \begin{tikzpicture}
        \begin{feynman}
        \vertex (a1) at (-1.5,1);
        \vertex (a2) at (1.5,1);
        \vertex (b1) at (-1.5,-1);
        \vertex (b2) at (1.5,-1);
        
        \vertex (m1) at (-0,1);  
        \vertex (m2) at (-0,-1); 

        \diagram* {
            (a1) -- [plain] (m1) -- [plain] (a2), 
            (b1) -- [plain] (m2) -- [plain] (b2), 
            (m1) -- [gluon] (m2), 
        };
        \end{feynman}
    \end{tikzpicture}
    \hspace{1.5cm}
    \begin{tikzpicture}
        \begin{feynman}
            \vertex (a) at (-1.1,0);
            \vertex (b) at (1.1,0);
            \vertex (c) at (0.7,0);
            \vertex (d) at (-0.7,0);
            \vertex (f) at (-1.1,-1.0);
             \vertex (g) at (1.1,-1.0);

             \diagram* {
                (a) -- [plain] (c),
                (d) -- [plain] (b),
                (c) -- [plain] (d),
                (c) -- [gluon, bend right=80] (d),
                (f) -- [plain] (g),
            };
        \end{feynman}
    \end{tikzpicture}
    \caption{One gluon exchange contribution to the energy of a $q\bar{q}$ configuration}
    \label{fig:oge}
\end{figure}
The individual quarks in a meson-core are considered so far to be experiencing the only force coming from the average effective potential $U(r)$ in Eq.\ref{eq:pot}. All that remains inside the meson-core is the hopefully weak one-gluon exchange as depicted in Fig~\ref{fig:oge}. The interaction Lagrangian density for the OGE can be expressed as,
\begin{equation}
    {\cal L}_I^g=\sum_a J_i^{\mu a}(x)A_\mu^a(x)
\end{equation}
where $A_\mu^a(x)$ are the vector-gluon fields and $J_i^{\mu a}(x)$ is the ith-quark colour-current. Since at small distances the quarks should be almost free, it is reasonable to calculate the shift in the energy of the meson-core (arising out of the quark interaction energy due to its coupling to the coloured gluons) using a first order perturbation theory. Then these contributions to the mass of meson due to the relevant diagram can be written as the sum of color-electric and color-magnetic energy shifts\\

\begin{equation}
    (\Delta E_M)_g=(\Delta E_M)_g^{\cal E}+(\Delta E_M)_g^{\cal M}
\end{equation}
where,
\begin{eqnarray}
    (\Delta E_M)_g^{\cal E}=&&\alpha_s \sum_{i,j}\; \Biggl<\sum_a \lambda_i^a\lambda_j^a \Biggr> \; \frac{1}{\sqrt{\pi}R_{ij}}\nonumber\\
    &&\times \; \bigg(1-\frac{\alpha_i+\alpha_j}{R_{ij}^2}+\frac{3\alpha_i\alpha_j}{R_{ij}^4}\bigg),
\end{eqnarray}
and
    \begin{eqnarray}
         (\Delta E_M)_g^{\cal M}=&&\alpha_s \sum_{i<j} \; \Biggl<\sum_a \lambda_i^a\lambda_j^a \sigma_i \sigma_j \Biggr> \; \frac{256}{9\sqrt{\pi}}\nonumber\\
    &&\times \; \bigg(\frac{1}{(3E'_i+m'_i)(3E'_j+m'_j)}\frac{1}{R_{ij}^3}\bigg)
    \end{eqnarray}
   
Here, 
\begin{eqnarray}
    R_{ij}^2=&&3 \; \Bigg(\frac{1}{E^{'2}_i-m^{'2}_i}+\frac{1}{E^{'2}_j-m^{'2}_j}\Bigg)\nonumber\\
    \alpha_i=&&\frac{1}{\lambda_i}(3E'_i+m'_i)
\end{eqnarray}
$\lambda_i^a$ are the Gellmann SU(3) matrices and $\alpha_s$ is the quark-gluon coupling constant. To describe a wide energy scale which covered light, strange and heavy mesons requires an effective scale-dependent strong coupling constant~\cite{PhysRevD.51.6348,PhysRevD.60.116008,PhysRevD.62.094031} that cannot be obtained from the usual one-loop expression of the running coupling constant because it diverges when $Q \to \Lambda_{QCD}$. The freezing of the strong coupling constant at low energies studied in several theoretical approaches~\cite{PhysRevLett.79.1209,Badalian:1997de} has been used in different phenomenological models~\cite{Mattingly:1993ej,PhysRevD.65.016004}. The momentum-dependent quark–gluon coupling constant is frozen for each flavour sector. For this purpose one has to determine the typical momentum scale of each flavour sector that, as explained in~\cite{PhysRevD.47.3013}, can be assimilated to the reduced mass of the system. As a consequence, we use an effective scale-dependent strong coupling constant given by
\begin{equation}
\alpha_s(\mu)=\frac{\alpha_0}{ln(\frac{\mu^2+\mu_0^2}{\Lambda_0^2})}  
\label{eq:alphas}
\end{equation}
where $\mu$ is the reduced mass of the $q\bar{q}$ system and $\alpha_0$, $\mu_0$ and $\Lambda_0$ are determined as explained in~\cite{Vijande:2004he}. This equation gives rise to $\alpha_s\simeq 0.54$ for the light-quark sector, a value consistent with the one used in the study of the non-strange hadron phenomenology~\cite{PhysRevC.59.428,Julia-Diaz:2001pxb,Garcilazo:2001ck,Julia-Diaz:2001apa} and it also has an appropriate high $Q^2$ behaviour, $\alpha_s \simeq 0.127$ at the $Z^0$ mass~\cite{Davies:1997mg}. Therefore, we use Eq. \ref{eq:alphas} for $\alpha_s$ values for respective mesons to predict the mass spectrum.\\
Taking into account the specific quark flavor and spin configurations in various ground-state mesons and using the relations
\begin{eqnarray}
\biggl<\sum_a(\lambda_i^a)^2\biggr>=&&\frac{16}{3}\nonumber\\
    \biggl<\sum_a\lambda_i^a \lambda_j^a\biggr>_{i\neq j}=&&\frac{-16}{3}
\end{eqnarray}
one can write the general expression for energy correction due to OGE.
\subsection{Centre of mass momentum and the meson mass ($m_M$)}
In this shell-type relativistic independent quark model, the motion of individual quarks within the hadron core does not inherently result in a state with a well-defined total momentum, as required for a physically consistent hadron state. A similar issue arises in nuclear physics, particularly in the case
$^3He$, as well as in bag models, necessitating an appropriate resolution~\cite{PhysRevD.21.1975,PhysRev.89.1102,wong1975generator}. The energy contribution from the spurious center-of-mass motion must be accounted for as an additional correction to the hadron energy, obtained from the individual quark binding energy over and above the perturbative corrections discussed in~\ref{subsec:oge}. This correction follows the approach proposed by Wong and other researchers~\cite{PhysRevD.24.1416,Duck:1976fp,Duck:1978mn,PhysRevD.29.1035,eich1985static}, which has been thoroughly detailed in the earlier work~\cite{Barik:1986mq}.

In such an approach, the static meson-core state with core-centre at X is decomposed into components $\Phi(P)$ of plane-wave momentum eigen states as
\begin{equation}
    |M(X)\rangle_c=\int\frac{d^3P}{W_M(P)}exp(iP.X)\Phi_M(P)|M(P)\rangle
\end{equation}
The inverse relation is 
\begin{equation}
    |M(P)\rangle=\frac{1}{(2\pi)^3}\frac{W_M(P)}{\Phi_M(P)}\int d^3Xexp(-iP.X)|M(X)\rangle_c
\end{equation}
with the normalisation as follows,
\begin{equation}
    \langle M(P')|M(P)\rangle=(2\pi)^3 W_M(P)\delta(P-P')
\end{equation}
and $W_M(P)=2w_p$. The momentum profile function $\Phi_M(P)$ can be obtained as 
\begin{equation}
    \Phi_M^2(P)=\frac{W_M(P)}{(2\pi)^3}\tilde{I}_M(P)
\end{equation}
where $\tilde{I}_M(P)$ is the fourier-transform of the Hill-wheeler overlap function \cite{PhysRevD.24.1416}. For mesons in the present model,
\begin{eqnarray}
     \tilde{I}_M(P)=&&\Bigg(\frac{r_{0q}^2}{2\pi}\Bigg)^{3/2}exp(-P^2r_{0q}^2/2)(1-6C_q+15C_q^2)\nonumber\\
     &&\Bigg[1+\frac{P^2r_{0q}^2(2C_q-10C_q^2+C_q^2P^2r_{0q}^2)}{(1-6C_q+15C_q^2)}\Bigg]
\end{eqnarray}
when, $C_q=(E_q'-m_q')/6(3E_q'+m_q')$

Now we estimate the center-of-mass momentum $P$ as,
\begin{eqnarray}
    \langle P^2\rangle=&&\int d^3P\tilde{I}_M(P)P^2\nonumber\\
    =&&\sum_q\langle p^2 \rangle
    \label{eq:comm1}
\end{eqnarray}
Here $\langle p^2 \rangle$ is the average value of the square of the individual quark-momentum taken over $1S_{1/2}$ single-quark state and is given by
\begin{equation}
    \langle p^2 \rangle_q=\frac{(11E'_q+m'_q)(E_q^{2'}-m_q^{2'})}{6(3E'_q+m'_q)}
    \label{eq:comm2}
\end{equation}

Thus we find that the zeroth order energy $E_M^0$ in Eq.\ref{eq:ezerom} for a ground-state meson $M$ arising out of the binding energies of the constituent quark and antiquark confined independently by a phenomenological average potential $U(r)$ must be corrected for the energy shifts due to the residual quark-gloun interactions discussed in~\ref{subsec:oge}. This would give the total energy of the $(q_i\bar{q}_j)$-system  in its ground state, as
\begin{equation}
    E_M=E_M^0+[(\Delta E_M)_g^{\cal E}+(\Delta E_M)_g^{\cal M}]
\end{equation}
Finally, taking into account the centre-of-mass motion of the $(q_i\bar{q_j})$-system with the center-of-mass momentum $P$ given as in \ref{eq:comm1} and \ref{eq:comm2}, one can obtain the physical mass of $(q_i\bar{q_j})$-meson in its ground state as,
\begin{equation}
    m_M=[E^2_M-\langle P^2\rangle_M]^{1/2}
    \label{eq:mmass}
\end{equation}
Since our main objective in this work is to obtain the masses of the $(q_i\bar{q_j})$-system, therefore Eq.~\ref{eq:mmass} is used to determine the masses of the kaons, charm and $b$-mesons as well as quarkonium.

\section{Results \& Discussions}
\label{sec:Res}
By accounting for relativistic effects, single-gluon exchange corrections, and center-of-mass motion corrections, we fine-tune the potential parameters and quark masses using experimental meson masses as benchmarks. For the quantitative evaluation of Eq.~\ref{eq:mmass}, it requires to determine the potential parameters ($a,V_0$) and quark masses as a first step. We consider these parameters as flavor-independent but treat them differently across a wide range of mesons according to their mass scale, which influences the confining strength. We differentiate the parameters among open flavors: kaons, charm, b-meson and also for quarkonium. 
The different set of ($a,V_0$), the quark binding energy $E_q$ and the results of the meson mass are mentioned in the Table~\ref{tab:my_label}.

To describe full spectrum of $S$-wave meson state (kaons, charm, b-meson), an effective scale dependent strong coupling constant must be incorporated. This parameter appears in the expression for the OGE correction and is taken into account in our calculations. We use the effective scale-dependent strong coupling constant as described in Eq.~\ref{eq:alphas}. This equation gives rise to $\alpha_s\simeq 0.54$ for the light-quark sector~\cite{Vijande:2004he}, which is consistent with that of used in the study of non-strange hadron spectroscopy~\cite{Barik:1987zb} in the framework of RIQ model. 

For predicting the meson mass we have taken the quark masses (in $GeV$) as,
\begin{eqnarray}
    m_{u/d}=0.305,\; m_s=0.510, \; m_c=1.500,\; m_b=4.700,\nonumber
\end{eqnarray}
 which align closely with those in constituent quark model~\cite{Vijande:2004he,Capelo-Astudillo:2025fnp}. Using this set of constituent quark masses, we fit the $(a,V_0)$ values for the masses of kaons and $\phi$ to reproduce those of experimental values and are obtained as $(a=0.374,V_0=-0.42)$, respectively. For charmonium and charm meson we choose $a=0.128\text{ and } V_0=-0.5$, which gives the best fit value. Then for the mesons with heavy quarks '$b$', we have taken $a\; \text{and}\; V_0$ values as 0.15 and -0.5, so that the Eq.~\ref{eq:mmass} yields $b$-meson masses which are comparable to experimental physical masses. For a comprehensive comparison, we present the meson masses obtained from our model alongside those predicted by the chiral quark model(QM)~\cite{He:2023gqh}. The corresponding experimental meson masses are also compared and included in Table~\ref{tab:my_label} for reference.
\begin{table}[!hbt]
    \centering
    \begin{tabular}{|cccccccc|}
    \hline
         \hline \multirow{2}{*}{Input parameters}&\multicolumn{3}{|c}{$a=0.374$} &\multicolumn{4}{c|}{$E_{u/d}=522.7$}\\
          &\multicolumn{3}{|c}{$V_0$=-0.42} &\multicolumn{4}{c|}{$E_s=632.7$}\\
          \hline \multicolumn{2}{|c}{Meson}& \multicolumn{2}{c}{Our work}& \multicolumn{2}{c}{Chiral QM}&\multicolumn{2}{c|}{Exp.}\\
          \hline \multicolumn{2}{|c}{$K$}& \multicolumn{2}{c}{490.5}& \multicolumn{2}{c}{503.3}&\multicolumn{2}{c|}{$493.67 \pm 0.015$}\\
           \multicolumn{2}{|c}{$K^*$}& \multicolumn{2}{c}{920.5}& \multicolumn{2}{c}{909.8}&\multicolumn{2}{c|}{$891.67 \pm 0.26$}\\
            \multicolumn{2}{|c}{$\phi$}& \multicolumn{2}{c}{1025}& \multicolumn{2}{c}{1027.6}&\multicolumn{2}{c|}{$1019.4 \pm 0.016$}\\
          \hline
          \hline \multirow{2}{*}{Input parameters}&\multicolumn{3}{|c}{$a=0.128$} &\multicolumn{2}{c}{$E_{u/d}=817.8$}&\multicolumn{2}{c|}{$E_c=1608.7$}\\
          &\multicolumn{3}{|c}{$V_0$=-0.5} &\multicolumn{2}{c}{$E_s=905.3$}& \multicolumn{2}{c|}{}\\
          \hline \multicolumn{2}{|c}{Meson}& \multicolumn{2}{c}{Our work}& \multicolumn{2}{c}{Chiral QM}&\multicolumn{2}{c|}{Exp.}\\
          \hline \multicolumn{2}{|c}{$D$}& \multicolumn{2}{c}{1848.3}& \multicolumn{2}{c}{1869.4}&\multicolumn{2}{c|}{$1864.8 \pm 0.05$}\\
           \multicolumn{2}{|c}{$D^*$}& \multicolumn{2}{c}{2191.0}& \multicolumn{2}{c}{2005}&\multicolumn{2}{c|}{$2006.8 \pm 0.05$}\\
           \multicolumn{2}{|c}{$D_s^{\pm}$}& \multicolumn{2}{c}{1958.9}& \multicolumn{2}{c}{1963}&\multicolumn{2}{c|}{$1968.3 \pm 0.07$}\\    
           \multicolumn{2}{|c}{$D_s^{\pm*}$}& \multicolumn{2}{c}{2240.2}& \multicolumn{2}{c}{2103.8}&\multicolumn{2}{c|}{$2112.2 \pm 0.4$}\\
           \multicolumn{2}{|c}{$\eta_c$}& \multicolumn{2}{c}{2765.8}& \multicolumn{2}{c}{2981.9}&\multicolumn{2}{c|}{$2984.1 \pm 0.4$}\\
           \multicolumn{2}{|c}{$J/\psi$}& \multicolumn{2}{c}{2929.9}& \multicolumn{2}{c}{3078}&\multicolumn{2}{c|}{$3096.9 \pm 0.006$}\\    
          \hline
          \hline \multirow{2}{*}{Input parameters}&\multicolumn{3}{|c}{$a=0.15$} &\multicolumn{2}{c}{$E_{u/d}=874.7$}&\multicolumn{2}{c|}{$E_c=1654.2$}\\
          &\multicolumn{3}{|c}{$V_0$=-0.5} &\multicolumn{2}{c}{$E_s=966.2$}& \multicolumn{2}{c|}{$E_b=4581.3$}\\
         \hline \multicolumn{2}{|c}{Meson}& \multicolumn{2}{c}{Our work}& \multicolumn{2}{c}{Chiral QM}&\multicolumn{2}{c|}{Exp.}\\
          \hline \multicolumn{2}{|c}{$B$}& \multicolumn{2}{c}{5301.8}& \multicolumn{2}{c}{5277.6}&\multicolumn{2}{c|}{$5279.7 \pm 0.08$}\\
           \multicolumn{2}{|c}{$B^*$}& \multicolumn{2}{c}{5448.8}& \multicolumn{2}{c}{5321.9}&\multicolumn{2}{c|}{$5324.75\pm0.20$}\\
           \multicolumn{2}{|c}{$B_s^0$}& \multicolumn{2}{c}{5346.1}& \multicolumn{2}{c}{5358.9}&\multicolumn{2}{c|}{$5366.9 \pm 0.10$}\\    
           \multicolumn{2}{|c}{$B_s^{0*}$}& \multicolumn{2}{c}{5469.3}& \multicolumn{2}{c}{5406.2}&\multicolumn{2}{c|}{$5415.4 \pm 1.4$}\\
           \multicolumn{2}{|c}{$B_c^\pm$}& \multicolumn{2}{c}{5976.3}& \multicolumn{2}{c}{6263.5}&\multicolumn{2}{c|}{$6274.4 \pm 0.32$}\\
           \multicolumn{2}{|c}{$B_c^{\pm*}$}& \multicolumn{2}{c}{6054}& \multicolumn{2}{c}{$-$}&\multicolumn{2}{c|}{$-$}\\  
           \multicolumn{2}{|c}{$\eta_b$}& \multicolumn{2}{c}{8936.5}& \multicolumn{2}{c}{ 9417.8}&\multicolumn{2}{c|}{$9398.7 \pm 2.0$}\\
           \multicolumn{2}{|c}{$\Upsilon$}& \multicolumn{2}{c}{8974.1}& \multicolumn{2}{c}{ 9448.2}&\multicolumn{2}{c|}{$9460.4 \pm 0.10$}\\    
          \hline
          \hline
    \end{tabular}
    \caption{Comparison of physical masses (in $MeV$) of ground state mesons with the experimental values ~\cite{ParticleDataGroup:2024cfk} and chiral QM~\cite{He:2023gqh}. The different set of potential parameters ($a$ $GeV^3$, $V_0$ $GeV$) and quark binding energy ($E_q$ $MeV$) are also given.}
    \label{tab:my_label}
\end{table}

In our predicted results, the mass difference between the spin-singlet $^1S_0$ and spin-triplet $^3S_1$ states of a meson leads to hyperfine mass splitting. The splitting is positive for spin-1 and negative for spin-0, meaning that the spin-1 meson is heavier - as can be seen in our predicted table. The variation in mass depending on the alignment of quark spins arises from the spin-spin interaction originating from the OGE. This provides insight into QCD dynamics, particularly the short-range behavior, and also helps to calibrate quark models and effective potentials. From the table, it is evident that the hyperfine splitting decreases with increasing meson mass, even though the system becomes more tightly bound. This trend is consistent with experimental values. The behavior matches expectations from the spin-spin interaction term, which becomes increasingly suppressed for heavier quark masses.

These calculations serve as a foundation for exploring higher excited states within each meson family, aligning with experimental indications of their abundance. Additionally, this framework offers a starting point for investigating exotic states within the relativistic independent quark model.
\section{Summarry and Concluding remarks}
\label{sec:Sum&con}
In the foregoing section, we adopt an equally mixed scalar-vector harmonic potential model to represent the non-perturbative multigluon interactions responsible for quark confinement in hadrons. This confinement is phenomenologically described by an effective potential. We incorporate the residual interactions into this quark confinement, which arises from quark-gluon coupling due to single-gluon exchange. These interactions are treated as low-order perturbations and include the spin-dependent terms which differentiates the masses between pseudoscalar and vector mesons, thereby reproducing hyperfine mass splitting. Furthermore, by accounting for the effects of the
center-of-mass motion, the model’s effectiveness is demonstrated through the fitting of quark masses and potential parameters $(a,V_0)$ to predict the spectrum of mesons across different categories of $S$-wave states: kaons, charm, and $b$-mesons. Additionally, an effective scale-dependent strong coupling constant is also introduced to refine calculations within RIQ model. The results closely reproduce experimental meson masses, providing a strong foundation for further studies on higher excited states and exotic mesons within this model framework.

\section*{Acknowledgements}
We sincerely express our deep gratitude to Professors N. Barik, P. C. Dash, and S. Kar for their valuable discussions and significant contributions in the development of the RIQ model framework. Additionally, we acknowledge Prof. Subhasis Basak for insightful discussions. We are also grateful to NISER, Department of Atomic Energy, India, for their financial support.

\bibliographystyle{unsrt} 
\bibliography{bibliography}






\end{document}